# Narrowband vacuum ultraviolet light via cooperative Raman scattering in dual-pumped gas-filled photonic crystal fiber

Rinat Tyumenev,[1*] Philip St.J. Russell[1,2] and David Novoa[1]

[1]Max Planck Institute for the Science of Light, Staudtstraße 2, 91058 Erlangen, Germany
[2]Department of Physics, Friedrich-Alexander-Universität, Staudtstrasse 2, 91058 Erlangen, Germany
*Corresponding author: rinat.tyumenev@mpl.mpg.de

**Many fields such as bio-spectroscopy and photochemistry often require sources of vacuum ultraviolet (VUV) pulses featuring a narrow linewidth and tunable over a wide frequency range. However, the majority of available VUV light sources do not simultaneously fulfill those two requirements, and few if any are truly compact, cost-effective and easy to use by non-specialists. Here we introduce a novel approach that goes a long way to meeting this challenge. It is based on hydrogen-filled hollow-core photonic crystal fiber pumped simultaneously by two spectrally distant pulses. Stimulated Raman scattering enables the generation of coherence waves of collective molecular motion in the gas, which together with careful dispersion engineering and control over the modal content of the pump light, facilitates cooperation between the two separate Raman combs, resulting in a spectrum that reaches deep into the VUV. Using this system, we demonstrate the generation of a dual Raman comb of narrowband lines extending down to 141 nm using only 100 mW of input power delivered by a commercial solid-state laser. The approach may enable access to tunable VUV light to any laboratory and therefore boost progress in many research areas across multiple disciplines.**

## Introduction

Modern bio-spectroscopy and chemistry in the vacuum ultraviolet (VUV, wavelength range from 100-200 nm) has recently emerged as a field of great interest since it gives access to outer-shell electronic transitions of proteins and amino-acids [1, 2]. These applications require sources of coherent, narrowband VUV light capable of covering a broad frequency range while offering high spectral brightness. The ideal solution would also be compact, cost-effective and easy to use by non-specialists in laser science. Apart from building-scale synchrotron facilities, several types of VUV laser are currently available, but few if any combine narrow linewidth and wide-band tunability. The most common example is the excimer laser, which relies on direct electronic transitions for lasing and is therefore inherently narrowband. These lasers are also bulky, fixed wavelength and have poor beam quality. Common alternatives based on frequency conversion in nonlinear crystals can be relatively efficient in the deep ultraviolet (DUV, wavelength range from 200-300 nm) [3]. These systems are however not widely tunable due to the strong material dispersion of most glasses and crystals in the ultraviolet, and furthermore absorption and photo-induced damage makes spectral extension below 180 nm difficult [4].

A promising alternative that is free of most of these limitations is gas-filled hollow-core photonic crystal fiber (HC-PCF) pumped by ultrashort near-infrared pulses. Although nonlinear effects such as self-compression and dispersive-wave emission make it possible to cover the whole DUV and a significant part of the VUV [5], the resulting spectra are relatively broadband and therefore low in spectral intensity, and moreover an ultrafast pump laser is required, making the system not easily portable and thus less attractive for many end-users.

Here we report the use of hydrogen-filled HC-PCF, pumped simultaneously by narrowband visible and UV pulses, to generate two vibrational Raman combs that through cooperative molecular modulation interact to produce a spectrum that reaches to 2.120 PHz (141.4 nm) in the VUV—much further than either of the two combs alone. This is made possible by deliberate excitation of higher-order guided modes that facilitate the onset of intermodal stimulated Raman scattering (SRS). A further attractive feature of this VUV source is the simple air-cooled solid-state pump laser (foot-print less than 0.02 $m^2$). The result is a highly compact (~0.5 $m^2$) and reliable light source that potentially will allow easy access to narrowband VUV light in any laboratory.

## Phase-matched Raman up-conversion

Above the SRS threshold, a narrowband pump pulse undergoes conversion to a frequency-down-shifted Stokes signal, accompanied by excitation of a coherence wave (Cw) of synchronous molecular motion (a "nonlinear hologram") that stimulates further growth of the Stokes signal. Within its coherence lifetime, this "nonlinear hologram" can also be used to frequency up-shift pump light (or indeed any signal) by the Raman frequency if the dephasing $\Delta\beta$ between the Cw and the spatio-temporal beat-note between the signals is small. Gas-filled HC-PCF not only strongly enhances the SRS gain [6], but also uniquely permits collinear phase matching over many 100s of THz through its pressure-tunable *S*-shaped dispersion profile around the zero-dispersion point (ZDP) [7] (Fig. 1). This arises from the balance between the normal dispersion of the gas and the anomalous geometrical dispersion of the modes of the evacuated hollow core (See Fig. 1).

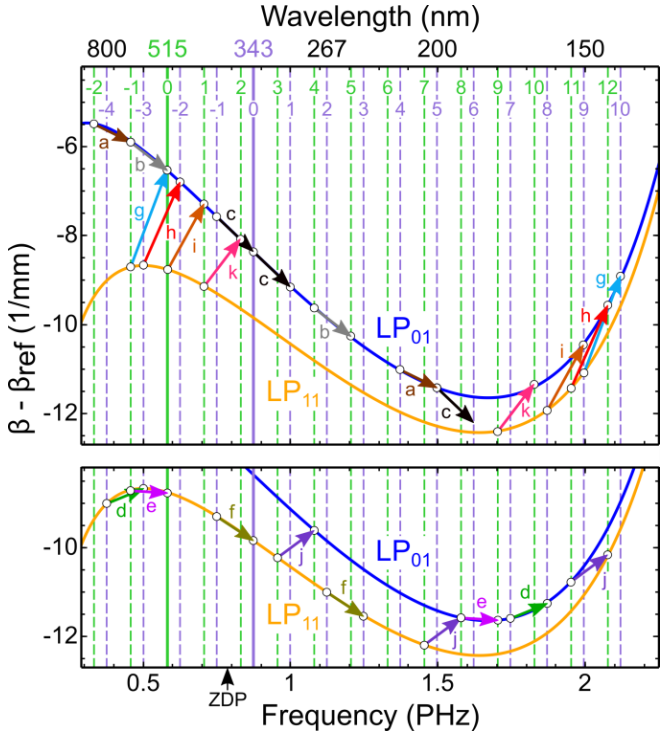

Fig. 1. Illustration of how intramodal and intermodal Raman Cws (marked by colour-coded arrows labelled a-k) in the vicinity of the two pump wavelengths (515 and 343 nm, marked by full vertical lines) can cause cascaded phase-matched up-conversion all the way to the vacuum UV. The dispersion curves are calculated for a kagomé-PCF (core diameter 22 μm) filled with 2.7 bar of $H_2$. To bring out the weak frequency dependence of the modal refractive indices $n$, the quantity $\beta - \beta_{ref}$ is plotted, where $\beta_{ref} = \omega\, n_{ref}/c$ and $n_{ref} = 1.0007655$. The group velocity dispersion is zero at 0.786 PHz for the $LP_{01}$ mode (marked with an arrow) and 0.965 PHz for the $LP_{11}$ mode. For clarity, given the many different Cws, the diagram is repeated twice. The numbers on the vertical dashed lines are the Raman orders for each comb. If the tip of a Cw arrow does not join the dispersion curve the up-conversion process is dephased. For example, the black arrow c in the upper plot at ~1.5 PHz has an intramodal dephasing length of ~5 mm (0.3 mm$^{-1}$ on the vertical axis corresponds to a dephasing length of 5 cm). Intermodal Cws play a major role in providing phase-matched up-conversion deep into the VUV.

Using hydrogen as a filling gas has many advantages: it has the highest Raman gain and frequency shift (~125 THz for the fundamental vibrational transition) of all gases and is transparent down to ~120 nm [8]. Combined with UV-enhanced SRS gain and ultralong interaction lengths, these features make $H_2$-filled HC-PCF very attractive for the generation of VUV light. Note that the $H_2$ dispersion becomes stronger as the electronic transitions are approached, making it more challenging to minimize the dephasing between the Cws (depicted by arrows in Fig. 1) and the spatio-temporal beat-note in the deep and vacuum UV.

Defining $\Delta\beta_{ij} = \beta_i(\omega) - \beta_j(\omega - \omega_R)$, where $i$ and $j$ can be any combination of 01 and 11, a Cw generated at frequency $\omega_P$ can be used for phase-matched up-conversion of a signal at $\omega$ provided the phase-matching condition $|\Delta\beta(\omega + \omega_R) - \Delta\beta(\omega_P)| < \pi/L$ is satisfied, where $L$ is the interaction length. Note that selection rules dictate that a Cw created by intermodal $LP_{01}$–$LP_{11}$ SRS cannot be used for intramodal conversion, and vice-versa.

For single-frequency pumping sufficiently close to the $LP_{01}$ ZDP (0.786 PHz in our case), phase matching can be fulfilled over a broad wavelength range and cascaded anti-Stokes bands generated down to ~200 nm [7, 9]. Although by appropriate PCF design the ZDP can be shifted into the DUV, permitting generation of VUV light, this requires access to a DUV pump laser [10].

Here we report that using UV and visible pumps, in combination with intramodal and intermodal Cws, offers multiple possibilities for close-to-phase-matched cascaded generation of narrow-band anti-Stokes signals (Fig. 1) reaching as far as 141.4 nm in the VUV.

**Experimental system**

A 50-cm-long kagomé-style HC-PCF (see inset in Fig. 2) with 22 μm core diameter and ~96 nm core-wall thickness was used. For these parameters the loss-band caused by anti-crossing between the core mode and the first core-wall resonance [11, 12] appears at ~225 nm. These parameters ensure optimal broadband guidance from the near-infrared to the VUV [13]. The fiber was mounted between two gas cells equipped with $MgF_2$ windows and connected through a metal tube. This allowed the fiber to be evacuated or filled with $H_2$ to a controllable pressure.

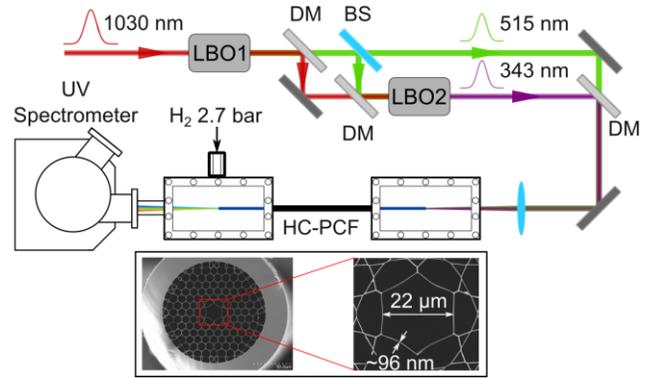

Fig. 2. Sketch of the experimental set-up. Dichroic mirrors and beam splitter are marked as DM and BS correspondingly. Beam splitter reflects 60% of 515 nm pump. LBO crystals for second and third-harmonic generation are marked as LBO1 and LBO2 accordingly. The inset at the bottom of the figure shows a scanning electron micrograph of the VUV-guiding HC-PCF employed.

The fiber was pumped by linearly-polarized, second (visible, 515 nm) and third (UV, 343 nm) harmonics of a commercial Q-switched laser (1030 nm, 1.5 ns pulses, 2 kHz repetition rate), generated in lithium triborate (LBO) crystals (see Fig. 2). A delay line was used to adjust the temporal overlap between the UV and visible pulses, and both pumps were then combined at a dichroic mirror and coupled simultaneously into the fiber using a thin lens. The pulse energies just before launching into the fiber were 21 µJ at 343 nm (70% transmission) and 38 µJ at 515 nm (80% transmission), and the output light was coupled to an evacuated VUV spectrometer via an airtight connection.

## Experimental Results

The main experimental results, obtained for a pressure of 2.7 bar (an optimal compromise between dispersion and Raman gain), are shown in Fig. 3. The VUV spectra were calibrated above 1.4 PHz by taking account of the frequency-dependent response of the detector, the diffraction efficiency and reflectivity of the grating, and the transmission of the $MgF_2$ output window. We also estimated the fraction of power blocked by the input spectrometer slit by numerically propagating both $LP_{01}$ and $LP_{11}$-like modes from the fiber endface and integrating their modal intensities over the 150 µm wide slit. The resulting spectral powers are plotted in Fig. 3(d). The overall conversion efficiency of pump energy to the 10$^{th}$ anti-Stokes at ~141.4 nm is ~10$^{-7}$. Based on the expected ~1 GHz linewidth of the Raman bands [14], we estimate the spectral power densities of the generated signals to be ~1 µW/GHz at ~185 nm and ~10 nW/GHz at ~141.4 nm, much higher than those previously reported by dispersive wave generation in HC-PCFs pumped by ultrashort pulses [5]. Note, however, that the limited resolution of the VUV spectrometer (~10 THz) prevented experimental confirmation of these values.

## Discussion

When pumped by each of the two colors separately (panels (a)–(b) in Fig. 3), the vibrational Raman comb lines extend across the whole DUV, cutting off at the VUV edge (~1.5 PHz). In the case of UV pumping, although the intramodal Cws (black arrows labelled c in Fig. 1) generated by SRS can be efficiently used for phase-matched anti-Stokes generation in the DUV, they become extremely dephased for intramodal up-conversion in the VUV (arrow c in the vicinity of 1.5 PHz).

For extension beyond 2 PHz, however, intermodal phase matching and single-frequency pumping are insufficient, since either some of the coherence waves needed for phase-matched cascaded up-conversion are missing, or there is no signal in some of the lower frequency comb lines.

By judicious choice of pump frequencies, chosen to optimally exploit the dispersion landscape, these limitations can be overcome, because the Cws generated by pump light at 515 and 343 nm in a mixture of $LP_{01}$ and $LP_{11}$ modes turn out to be sufficient to provide an uninterrupted chain of up-conversions reaching deep into the VUV. This may be seen in Fig. 1, where the intramodal Cws labelled (a-f) are sufficient to reach 1.9 PHz, while the intermodal CWs labelled (g-k) provide the additional momentum required to reach 2.120 PHz, along the upward-climbing dispersion curves.

### Numerical modeling of dual Raman comb dynamics

The simple picture presented above of cascaded phase-matched anti-Stokes generation cannot capture all the features of the up-conversion process, which is highly complex, containing multiple pathways. To gain further insight, therefore, we modelled the system using multimode coupled Maxwell-Bloch equations [13], considering the overall coherence to be the superposition of the Cws created by the two combs separately. Fifteen bands (pump, 4 Stokes, 10 anti-Stokes, see Table 1) were included for each pump frequency. For the sake of simplicity and inspired by the experimental observations, we considered only the $LP_{01}$ and $LP_{11}$ modes and disregarded the effects of rotational SRS and fiber loss. The best agreement with the experiments was obtained for pulse energies of 31 µJ at 515 nm and 21 µJ of 343 nm, and a modal content of 65%/35% $LP_{01}/LP_{11}$. The left-hand column in Fig. 4(a) shows the powers generated in the UV comb lines, and the right-hand column the same information but for the visible comb lines. In the top row, the fiber is pumped by one color at a time, showing that, as in the experiment (Figs. 3(a)–(b)), the simulated combs do not reach deep into the VUV (grey shaded area), except for a few very weak lines in the 343 nm comb.

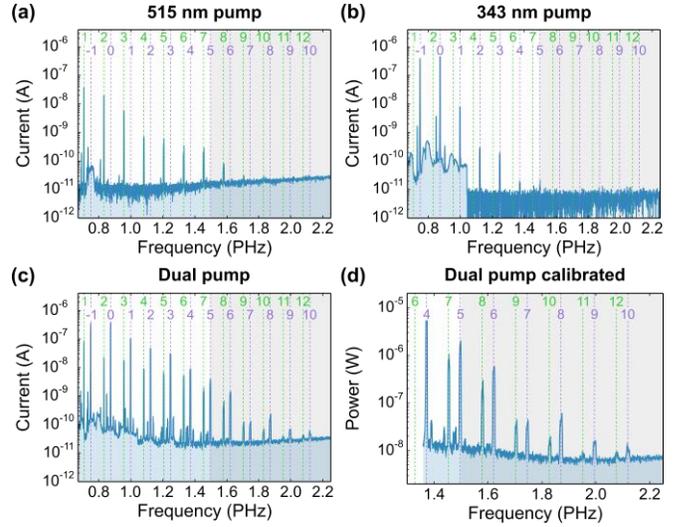

Fig. 3. Experimental spectra obtained in the fiber filled with 2.7 bar $H_2$. The different panels show the anti-Stokes side of the Raman combs generated when the fiber is pumped solely by (a) visible pulses, (b) UV pulses, and (c) both visible and UV pulses. Panels (a)–(c) show uncalibrated spectra (the y-axis represents the current measured by the photodiode in the VUV spectrometer). Panel (d) shows the calibrated power in the spectra shown in (c) above 1.4 PHz. Grey-shaded areas mark the VUV domain.

| Sideband order | UV comb | | Visible comb | |
| --- | --- | --- | --- | --- |
| | λ (nm) | ν (PHz) | λ (nm) | ν (PHz) |
| –4 | 797.8 | 0.376 | 3576.0 | 0.084 |
| –3 | 599.2 | 0.500 | 1438.0 | 0.208 |
| –2 | 479.8 | 0.625 | 900.3 | 0.333 |
| –1 | 400.0 | 0.749 | 655.2 | 0.457 |
| 0 | 343.0 | 0.874 | 515.0 | 0.582 |
| 1 | 300.2 | 0.999 | 424.2 | 0.707 |
| 2 | 266.9 | 1.123 | 360.6 | 0.831 |
| 3 | 240.3 | 1.248 | 313.6 | 0.956 |
| 4 | 218.5 | 1.372 | 277.5 | 1.080 |
| 5 | 200.3 | 1.497 | 248.8 | 1.205 |
| 6 | 184.9 | 1.621 | 225.5 | 1.329 |
| 7 | 171.7 | 1.746 | 206.2 | 1.451 |
| 8 | 160.3 | 1.871 | 190.0 | 1.579 |
| 9 | 150.3 | 1.995 | 176.0 | 1.703 |
| 10 | 141.4 | 2.120 | 164.0 | 1.828 |

Table 1. All sideband orders observed and included in the simulations. Stokes lines are indicated with a minus sign and the pump is the zero-order band in each comb. Both wavelengths λ and the corresponding frequencies ν are displayed. The shading marks the VUV lines.

In sharp contrast, for dual-pumping (middle row in Fig. 4(a)) the VUV extension of both combs is dramatically enhanced, reaching out to the shortest wavelengths observed in the experiment (~141.4 nm, 2.120 PHz) and confirming that cooperative molecular modulation is indeed key to generating narrowband VUV lines in gas-filled HC-PCF.

In addition, by comparing the experimental near-field mode profiles for selected DUV lines (see Fig. 4(b)) with the simulations, we see that the modelling also accurately predicts the modal content of each sideband, revealing for example that the UV comb is more dependent on the higher-order mode fraction than the visible comb. To further confirm this, we simulated the ideal case of a HC-PCF supporting only the fundamental mode (bottom row in Fig. 4(a)). By thus switching off intermodal SRS and launching all the pump energy in the $LP_{01}$-like mode, the spectral extent of the visible comb remains almost unaltered while the UV comb significantly shrinks, confirming the crucial role played by intermodal SRS in VUV light generation.

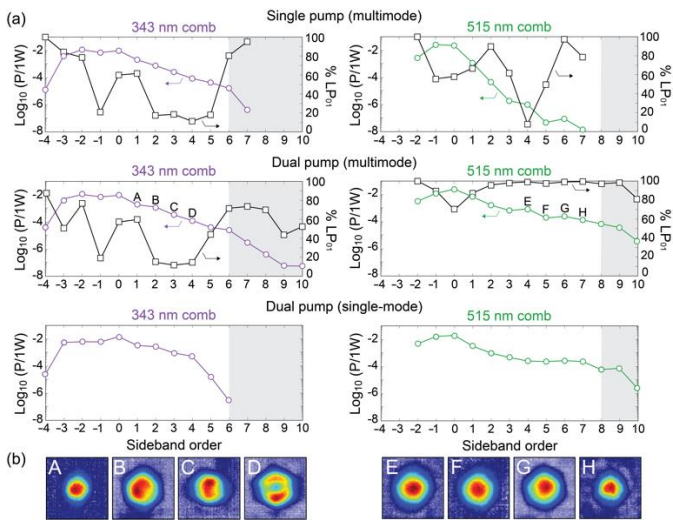

Fig. 4. (a) Numerically calculated power P (logarithmic scale) in the Raman comb lines (see Table 1) for 343 nm pumping (left column) and 515 nm pumping (right column) in a HC-PCF filled with 2.7 bar $H_2$. The top row is for one-color pumping, the middle row for dual pumping and the bottom row for dual pumping in an ideal single-mode HC-PCF, i.e., with intermodal nonlinear effects switched off. The right-hand axis in the upper two rows gives the proportion of the power in the $LP_{01}$-like mode. The grey-shaded areas mark the VUV domain. The 3rd and 4th Stokes bands of the 515 nm comb were artificially attenuated to account for the fact that they are not guided in the fiber used in the experiments. For further details on the parameters of the simulations, see text. (b) Experimental near-field mode profiles for the comb lines marked A-H in the middle row of (a).

## Conclusions

In conclusion, two non-overlapping Raman combs simultaneously generated in a $H_2$-filled HC-PCF dual-pumped by 525 nm and 343 nm light can exchange energy and increase their VUV content via cooperative molecular modulation. Experiments and simulations confirm that intermodal SRS plays a key role in extending the Raman combs beyond 2 PHz. The HC-PCF system provides narrowband VUV pulses down to 141 nm with spectral densities exceeding 10 nW/GHz using only 100 mW of input power delivered from a simple air-cooled compact laser. The efficiency of the system might potentially be further increased using pressure gradients or tapered HC-PCFs [15] to improve phase matching. Moreover, the average power in the VUV could be upscaled by using pump lasers with higher repetition rates [16].

We believe that the results may lead to the first truly portable solution for delivery of tunable VUV light in any laboratory, and may boost many research areas whose progress has been hampered by the lack of suitable turn-key table-top VUV sources.